# Real-time sensing with multiplexed optomechanical resonators

F. R. Lamberti[1], U. Palanchoke[1], T. P. J. Geurts[1], M. Gely[1], S. Regord[1], L. Banniard[1], M. Sansa[1], I. Favero[2], G. Jourdan[1], S. Hentz[1*]

1 Université Grenoble Alpes, CEA, LETI, 38000 Grenoble, France

2 Matériaux et Phénomènes Quantiques, CNRS UMR 7162, Université de Paris, 75013 Paris, France

*e-mail: sebastien.hentz@cea.fr

## Abstract:

Nanoelectromechanical resonators have been successfully used for a variety of sensing applications. Their extreme resolution comes from their small size at the cost of low capture area, making the "needle in a haystack" issue acute. This leads to poor instrument sensitivity and long analysis time. Moreover, electrical transductions are limited in frequency, which limits the achievable mechanical bandwidth again limiting throughput. Multiplexing a large number of high-frequency resonators appears as a solution, but this is complex with electrical transductions. We propose here a route to solve these issues, with a multiplexing scheme for very high frequency optomechanical resonators. We demonstrate the simultaneous frequency measurement of three silicon microdisks resonators fabricated through a Very Large Scale Integration process. The readout architecture is simple and does not degrade the sensing resolutions. This paves the way towards the realization of sensors for multi-parametric analysis, extremely low limit of detection and response time.

## Introduction:

Nanomechanical resonators have demonstrated their ability to perform chemical[1,2], mass[3,4], biological[5,6], force[7,8] and stiffness[9] sensing. Record limits of detection have been reached[10,11], and the possibility to fabricate and integrate them in a very large scale integration (VLSI) context has been demonstrated[12]. An increasing number of applications is reaching the market[13–16]. These achievements have been made possible by the extreme resolution of nanomechanical resonators, which itself stems from their miniature size. For many of these sensing applications though, reduced dimensions come at the cost of low throughput. This is particularly true for chemical, mass and biological applications where capture area is key in analysis time and system sensitivity (required minimum quantity or concentration at the system input). Both degrade with decreasing mechanical element size due to increased analyte diffusion time or interaction with nonactive sensor regions. Significant research efforts have been dedicated to solve this issue by multiplexing a number of resonators, with both external[17,18] and integrated[19–21] electrical transductions of motion. These solutions are all limited either by achievable frequency, speed, number of multiplexed devices, or fabrication complexity. Moreover, throughput is also ultimately limited by the mechanical resonator's bandwidth, which is primarily dictated by its resonance frequency[22]. Unfortunately, electrical transductions at the nanoscale are also limited in achievable frequency. Finally, the achievable mechanical sensing data rate, determined by resonator's bandwidth and number of resonators, is limited when using nanoscale electrical transductions.

In parallel with the developments of arrayed devices employing external or integrated electrical transductions, the field of cavity nano-optomechanics has advanced with great strides in the last decade. Long driven by fundamental studies[23,24], it has eventually approached technological maturity, building on the development of integrated photonics. Combined with significant assets such as extreme displacement sensitivity and virtually unlimited bandwidth in the transduction of motion, this maturity enabled demonstrations of inertial[25,26], chemical[27], biological[28], force[22], and mass[29] sensing. The first large scale fabrication capabilities in optomechanics were reported[29–31], adapting industrial-grade process flows from photonics to a silicon-on-insulator (SOI) platform enabling nanomechanical resonators. Large bandwidth being a prominent feature of datacom silicon photonics, with integrated modulators now reaching data rate in excess of 100 Gb/s, the vision naturally emerges to increase the data rate of nanomechanical sensing applications. This can be achieved by pushing up the frequency of nanomechanical devices, and by operating a large number of them in parallel using wavelength division multiplexing (WDM) techniques borrowed from the telecom industry. Preliminary steps in the direction of multiplexing in optomechanics were taken with beam[32] and membrane[33] mechanical resonators coupled to optical cavities addressed with a single bus waveguide. In absence of multiplexer/demultiplexer, a single continuous-wave laser was used and resonators were addressed one by one sequentially by tuning the laser wavelength. The mechanical frequencies employed in this work were low, typically in the 10 MHz range or below, greatly limiting the reachable throughput. A real-time optomechanical sensing experiment using several high-frequency resonators operated simultaneously by WDM, with a clear path towards scalability, thus remains to be demonstrated. Such demonstration is a mandatory step toward high data-rate mechanical sensing enhanced by photonics, but requires a controlled fabrication process, as well as a specific design and readout scheme that do not degrade the sensor's performance.

We report here a real-time sensing experiment performed with three very high-frequency (VHF) optomechanical resonators. We demonstrate a multiplexing scheme allowing simultaneously addressing and monitoring the individual mechanical resonance frequency of each resonator. The dimensions of resonators are finely tuned with our large-scale process, such that multiple optical fields of distinct wavelength can simultaneously probe the resonators' mechanical motion and carry the sensing information through a single bus waveguide. Demultiplexing is obtained thanks to an imposed difference in mechanical frequency amongst the three resonators. Each resonator motion is coherently driven using nearby electrodes and operated in closed loop, in order to individually monitor its mechanical resonance. We show that our multiplexing scheme does not degrade the resonator frequency stability – nor the limit of detection - and perform multiplexed sensing of an air flow in real-time conditions.

## Results:
**Optomechanical multiplexing for sensing.**
Contour mechanical modes of microdisks – or microrings - display high mechanical resonance frequency, high stiffness, small displacement and tiny effective mass. These features make them unique devices for, for instance, force sensing[22], as well as mass sensing in liquid in which they retain high performance[30,31,34]. The devices employed in this work are monocrystalline silicon disk resonators with diameter in the 20 μm range. They sustain optical whispering gallery modes (WGM) (Fig. 1a) as well as mechanical radial breathing modes (RBM) (Fig. 1b). Their mechanical motion is transduced as a modulation of the output optical intensity *via* the optomechanical interaction.

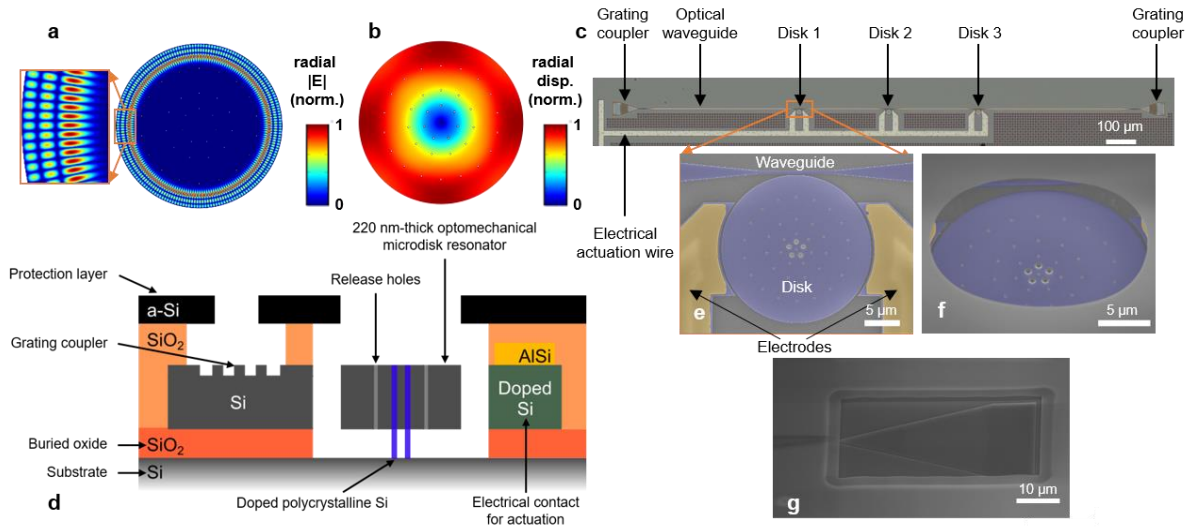

**Figure 1: Multiplexed monocrystalline silicon optomechanical disk resonators. a** Finite element simulation of an optical transverse electric whispering gallery mode at 1560 nm wavelength, radial number $p = 3$ and azimuthal number $m = 92$, for a 20 µm diameter microdisk. The color indicates the intensity of the electrical field in the radial direction. **b** Finite element simulation of the fundamental mechanical radial breathing mode at 253.12 MHz, for a 20 µm diameter microdisk. The model includes the anisotropy in the elastic properties of silicon. The color indicates the modulus of the radial displacement. **c** Optical microscope image. Three disks with respectively 20.12, 20.06 and 20 µm diameter are evanescently coupled to the same optical waveguide (with a 200 nm gap). Metal pads and leads allow the polarization of nearby (300 nm gap) electrodes for actuation. **d** Cross-section after fabrication. The 220 nm silicon-on-insulator (SOI) top layer is etched to pattern the optical grating couplers, waveguides, resonators and electrodes (see Methods). Each disk is supported by five highly p-doped poly-Si central pillars for anchoring and electrical polarization. The Si layer is highly doped locally below the AlSi leads for low contact resistance. A 200 nm amorphous silicon (a-Si) layer is deposited above a silicon oxide layer for protection, and etched open above the grating couplers and the resonators (not visible in panels c and e). The disks are released by HF vapor oxide etching thanks to release holes (visible in panels b, e and f) that both decrease the required etching time and suppress optical higher-order radial optical modes. **e** Zoom-in on a false-colored SEM image of a resonator, without a-Si protection, for clarity. The release holes and the poly-Si anchors are visible at the disk surface, with radii of 100 nm and 250 nm, respectively. **f and g** SEM images of the final resonator and a grating coupler with the a-Si protection.

Three disk resonators are evanescently coupled to a same bus optical waveguide, itself terminated by grating couplers for in and out on-chip light coupling[29] (Fig. 1c). The devices were fabricated with a recently developed optomechanical Very Large Scale Integration (VLSI) process[29], starting from 200mm SOI wafers (see Fig. 1d and Methods). The variable shape beam lithography used in this process in conjunction with carefully chosen silicon etching demonstrated low optical losses and optical modes with internal quality factor in excess of a million[35]. An amorphous silicon (a-Si) layer is added for protection from potential adsorbates and is etched open only on top of the resonators and of the grating couplers (see Figs. 1d, 1f and 1g). The limit of detection for resonant mechanical sensing is proportional to the measured mechanical frequency stability, itself inversely proportional to the signal-to-noise ratio (SNR) in a regime where additive noise dominates[36]. Moreover, coherent mechanical actuation is generally used to track the resonance frequency in real time with a feedback loop on the phase response. To this purpose, coherent mechanical drive is obtained with nearby electrodes and metallic leads. Fig. 1e is a scanning electron microscope (SEM) image showing all the elements surrounding a resonator in the absence of the a-Si protection layer. Furthermore, five p-doped polycrystalline Si anchors sustain the optomechanical disk resonator and allow grounding its voltage *via* the substrate for electrostatic actuation.

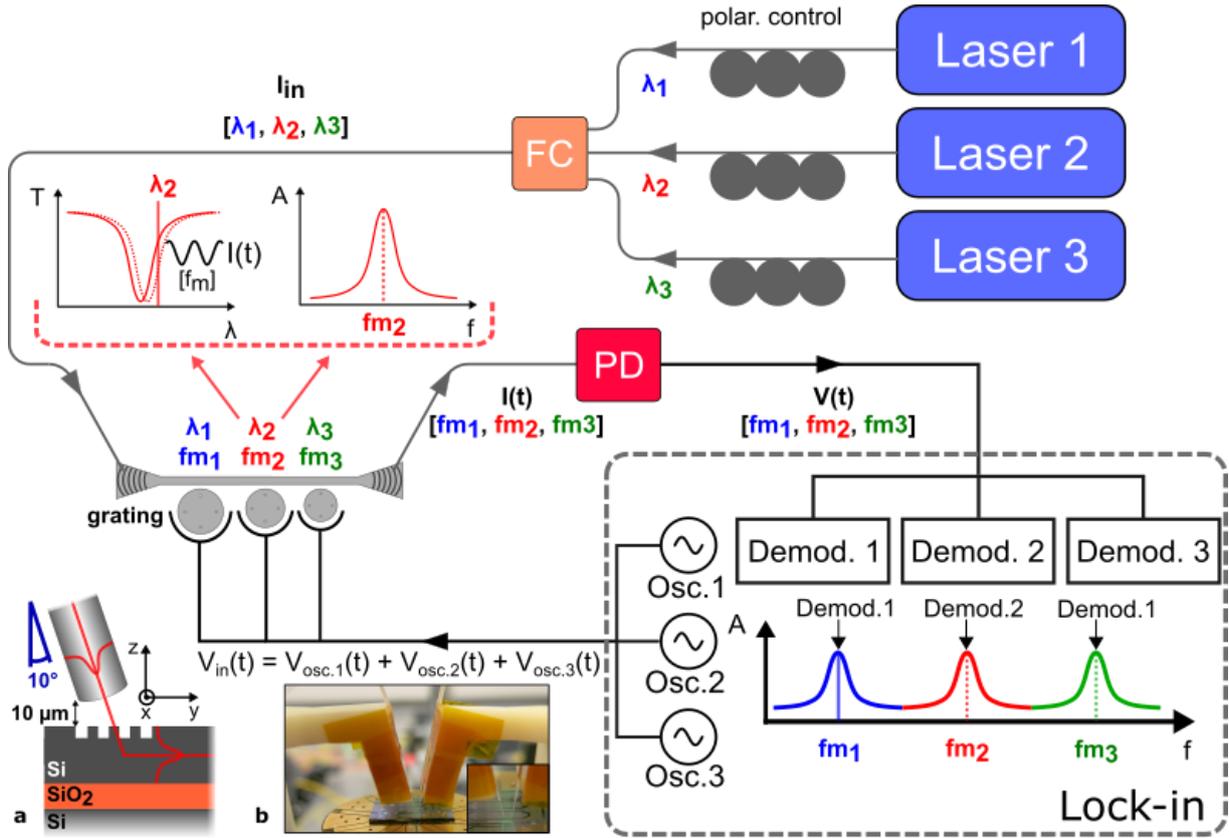

**Figure 2: Optomechanical multiplexing scheme.** Three continuous-wave lasers are coupled into the same optical fiber via a fiber optic coupler (FC, see Methods). Each laser wavelength is chosen to correspond to an optical resonance of a distinct disk. The injection and collection optical fibers are aligned ≈ 10 μm above the in and out grating couplers, using 3-axis piezoelectric translations stages and fiber holders having 10° angle to the vertical direction[37], as represented in **a**. The fibers are aligned above the silicon chip hosting the resonators, enabling to optically address the microdisks and collect the output optical signals as shown in **b**. The devices are driven at their mechanical resonance frequency with three signal generators of a lock in amplifier (LIA). As a result of the optomechanical coupling, the output optical intensity associated to each laser gets then modulated at the mechanical frequency of the corresponding disk, and the total sum of intensities, is collected by a photodiode (PD). The total PD output electrical signal is finally demodulated with three channels, each tuned to one particular drive frequency, enabling open- and closed loop operations for resonance frequency tracking.

Disks 1, 2 and 3 have a diameter of 20.12, 20.06 and 20 μm, leading to expected fundamental RBM frequencies fm$_1$ =251.66 MHz, fm$_2$ =252.39 MHz and fm$_3$ =253.12 MHz, respectively. This leads to a clear separation of the microdisks resonances for mechanical quality factors in the few thousands range. Expected WGM wavelengths are also distinct for each disk, with *ca* 3.3 nm separation for modes of identical radial and azimuthal numbers (see Supplementary Note 1). Each resonator, with loaded optical quality factors in the $10^5$ range, can then be addressed with a different excitation wavelength. Therefore, each disk $i$ corresponds to a mechanical frequency/optical wavelength doublet $(f_{mi}, \lambda_i)$. The electrostatic actuation is performed with a metallic lead that is common to all three disks.

The experimental set-up is described Fig. 2 (all measurements were performed in air). Three AC voltages (with frequencies set to fm$_1$/2, fm$_2$/2 and fm$_3$/2, respectively) are combined in the actuation lead to drive the mechanical resonators around their individual resonance frequency. Similarly, three optical fields, stemming from three distinct lasers, are combined into a single optical fiber with each optical wavelength set within one optical resonance of a distinct disk. One end of the optical fiber was cleaved, and aligned in front of a grating coupler with piezoelectric stages (see Fig. 2). In each disk, the mechanical motion modulates the resonance wavelength of the WGM, such that the intensity of each optical field is modulated at the mechanical frequency of the corresponding disk. The modulation amplitude increases with the $Q_{opt} C_r g_{om}$ product, where $Q_{opt}$ is the loaded quality factor of the considered optical mode, $C_r$ its optical contrast, and $g_{0m}$ the optomechanical coupling factor, calculated here at a reduction point of maximum displacement[29,38]. The total output optical signal is collected at the output grating by a second optical fiber connected to a photodiode. The information of the three disks' mechanical motion is then carried by an electrical signal modulated at $fm_1$, $fm_2$ and $fm_3$. It is subsequently demultiplexed by three distinct demodulators (see Methods).

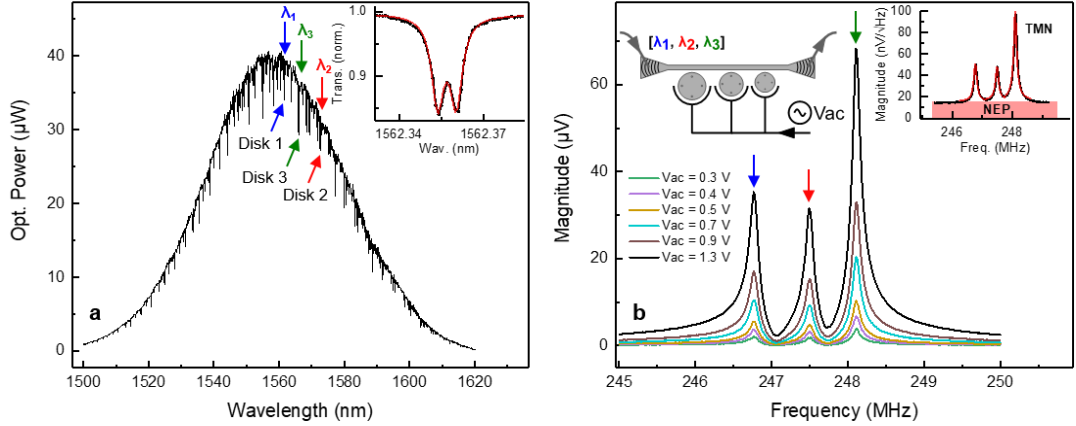

**Figure 3: Optical and mechanical responses of the multiplexed resonators. a** Optical power transmitted through the waveguide as a function of laser wavelength. The incident laser power is measured at the input of the coupling fiber, shown in Fig 2.a, and is here of 2mW. The red, blue and green arrows each indicate an optical mode of a distinct disk resonator. They are identified as the $(p = 3, m = 93)$, $(p = 3, m = 92)$ and $(p = 3, m = 92)$ modes, for Disk 1, 2 and 3, respectively. Inset: Low power normalized waveguide transmission for $(p = 3, m = 93)$ Disk 1's optical mode (16µW input laser power). A doublet is clearly visible and is fitted (red line) to determine the optical quality factors. **b** Mechanical response in the electrical domain as a function of drive frequency (see left inset), for drive amplitudes $Vac$ ranging from 0.3V up to 1.3V and 2 mW input laser power each (see Supplementary Fig. 5). The response obtained for Vac = 1.5 V is shown in Supplementary Note 3. Right inset: thermomechanical spectrum of the three disks (i.e. when Vac = 0V). The fits in red allow determining the mechanical resonance frequencies and quality factors, which are of 2533, 2534 and 2942 for Disk 1, 2 and 3, respectively. The noise floor is set by the photodetector noise equivalent power (NEP).

**Characterization of optical and mechanical responses.** To first characterize the optical response of the device, only one excitation laser is used (see Fig. 3a). Optimized grating couplers and waveguide geometry ensure efficient excitation only of transverse electric (TE) fields. The Gaussian broadband transmission response visible in Fig. 3a is typical of grating couplers[37]. We also obtain sharp WGM resonance lines corresponding to TE WGMs of different radial and azimuthal $(p, m)$ numbers[39]. The correspondence of observed resonances with the three disks can be unraveled by shining a red laser spot onto the top of each disk successively. This heats-up the targeted disk resonator and red-shifts its optical resonances through the thermo-optic effect, while leaving the other two disks unaffected (See Supplementary Fig. 4). Numerical simulations allow identifying the $(p, m)$ numbers for all the TE optical resonances belonging to the different disks (see Supplementary Note 1). The inset of Fig. 3a shows one of Disk 1 's optical resonance and is representative of the three disks response. A doublet splitting[40] is resolved thanks to low optical losses. Fitting such optical transmission spectrum yields loaded quality factors ranging from 2 to 3.5 $10^5$, with an optical contrast $Cr$ between 15 and 20% (see Supplementary Notes 2 and 4). The chosen 200 nm gap distance between the waveguide and disks is the result of a conservative choice (we estimate that the critical coupling regime $Cr = 100\%$ would be reached for a ≈ 100 nm gap).

Once optical characterization is performed, we select optical resonances to be used for mechanical measurements. They are chosen so that they present the same radial number for each disk, which leads to similar $g_{om}$, and high values for their optical contrasts and quality factors (see Supplementary

Notes 2 and 3). They correspond to the $(p=3, m=93)$, $(p=3, m=92)$ and $(p=3, m=92)$ (see Fig. 1a) modes for Disk 1, Disk 2 and Disk 3, respectively. Their spectral positions are indicated in Fig. 3a. The distinct wavelengths of the three lasers correspond to those of the selected optical modes, as indicated by the arrows in Fig. 3.a, and the drive voltages are set to zero (Vac = 0 in Fig. 3b). The thermomechanical noise spectra of the three resonators are simultaneously measured (inset of Fig. 3b). Our transduction and set-up prove to be sensitive enough to clearly resolve the Brownian motion of the three disks, with amplitudes of a factor 3.3 up to 6.7 above the noise floor. The latter is dominated by the noise-equivalent power of the photodiode (NEP). The theoretical Brownian motion amplitude at resonance in the mechanical domain is $\sqrt{\frac{4k_b T_0 Q_m}{m_{eff}\Omega_m^3}} \approx 0.3\ fm.\sqrt{Hz}^{-1}$. ~10 % discrepancies are expected for the three disks due to differences in diameter and mechanical quality factor. Yet, the ratio of the largest to the smallest noise peak amplitude in the electrical domain is around 2 (see right inset Fig. 3b). We attribute this in particular to variability in the values of the optomechanical coupling factors, which range from $\frac{g_{om}}{2\pi} = 12.3$ to 15.7 GHz.nm$^{-1}$, and in the contrast and quality factors of the disks's optical resonances (see Supplementary Note 3). The measured mechanical resonance frequencies are 246.78, 247.50 and 248.11 MHz respectively, showing a discrepancy of less than 2% with respect to the simulated values. By carrying out measurements on 12 different microdisks taken from different samples, we obtain a standard deviation for this parameter of 0.29 MHz, indicating a good reproducibility of our fabrication process. The mechanical quality factors range from 2500 up to 2900 (see Fig 3.b), translating into mechanical bandwidths in the 100 kHz range, which allows a clear separation of the peaks and a convenient demodulation.

The mechanical resonators are then driven with an increasing drive voltage. A single demodulation channel is used, with demodulation frequency swept around the three mechanical resonance frequencies (see Fig. 3b left inset). The motion amplitude scale like the square of the drive voltage, as expected (Fig. 3b). The maximum available voltage amplitude (1.5 V), while not sufficient to reach the non-linear mechanical threshold (see Supplementary Note 3), already yielded a typical SNR of ≈ 762 for the three disks, calculated for a 1Hz bandwidth and for a noise floor set by the thermomechanical noise. Moreover, the phase response displayed strong phase variations while sweeping over the resonances (see Supplementary Fig. 11), which demonstrates low background signal in the phase and facilitates closed-loop operation.

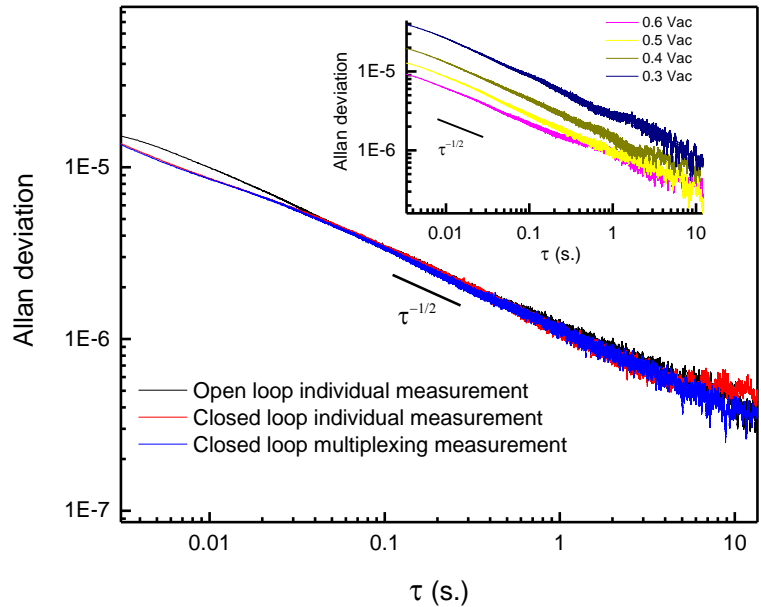

**Figure 4: Frequency stability measurements in individual and multiplexing operation.** Black and red lines: Open- and closed (respectively) loop Allan deviation for Disk 2 in individual operation: one laser (2mW laser power), one drive frequency (0.5 V amplitude) and one demodulation channel. Blue curve: Closed-loop Allan deviation for Disk 2 in multiplexing operation: three lasers (each one with an output power of 2 mW), three drive signals (each one with a drive amplitude of 0.5V) and three PLL channels. The excellent overlap demonstrates that the frequency stability is not affected by the multiplexing scheme. Inset: Open-loop Allan deviation of Disk 2 as a function of drive amplitude in individual operations. The Allan deviation scales like $\tau^{-1/2}$ and like $V_{ac}^{-2}$, which is consistent with a dominant additive white noise such as thermomechanical noise.

**Frequency stability in multiplexed operation.** The Allan deviation of each disk was measured in both open and phase-locked loop (see Supplementary Fig. 12) and compared between individual operation (one laser, one demodulation channel) and multiplexed operation (three lasers, three demodulation channels in parallel) for the same drive voltage. The result for Disk 2 is shown Fig. 4 with integration times τ spanning four orders of magnitude (the result is very similar for all disks, see Supplementary Fig. 13). Open and closed-loop Allan deviations in individual operation almost perfectly overlap, showing the innocuity of the phase locked loop (PLL) corrector to the eigenfrequency measurement, together with its ability to properly track fast frequency variations. The plot scales like $\tau^{-1/2}$ over the investigated time range. Moreover, the inset of Fig. 4 shows that the deviation is inversely proportional to the drive amplitude, and therefore to the square of the drive voltage. These are indications that in the individual disk operation, the frequency stability is limited by additive white noise, itself dominated by thermomechanical noise[36]. The Allan deviation obtained for Disk 2, while operating the three disks in closed-loop with our multiplexed scheme, is again almost identical to the one in individual operation. This proves that our multiplexing scheme does not degrade the frequency stability of our devices or the sensing limit of detection. The frequency stability reaches a value of the order of $10^{-6}$ for 1 second integration time, which is remarkable for devices vibrating in ambient pressure at ≈ 250 MHz.

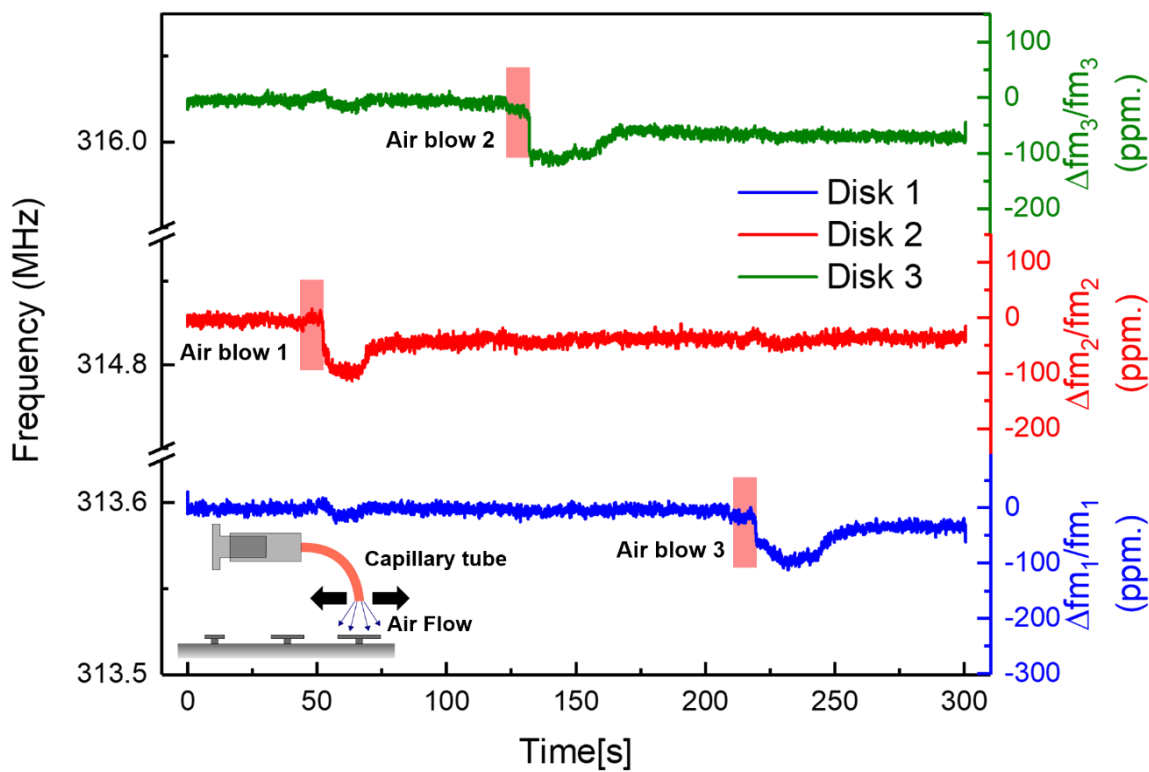

**Figure 5: Demonstration of real-time sensing in multiplexing operation.** A capillary tube connected to a syringe is positioned with micrometric stages ≈ 50 µm above a resonator. The air flow induced by pressing the syringe changes the mechanical resonance frequency of the resonator. The blue, red and green lines are the frequency timetraces of Disk 1, Disk 2 and Disk 3 respectively, with 0.2 s integration time. The red rectangles indicate the time intervals when the syringe piston starts and finishes moving.

**Multiplexed sensing demonstration.** We induce changes in mechanical frequency by flowing ambient air perpendicularly to a given disk surface with a capillary, similarly to measurements at the output of a gas chromatography column or in a chemical test bench[2,20,27,41] (see Fig. 5 and Supplementary Fig. 14). In parallel, we monitor the frequency of each disk simultaneously over time with our multiplexing scheme. For this experiment, a device with disks far apart from each other (1050 µm) was chosen so that the air flow is mainly sensed by only one disk. Disk 1, 2 and 3 diameters are here 16.12, 16.06 and 16µm respectively with ≈ 313.60, ≈ 314.83 and ≈ 316.02 MHz resonance frequency respectively. The capillary is first placed on top of the central resonator (Disk 2), then moved to Disk 3, followed by Disk 1. The ~30 KHz sharp decrease (~100 ppm) in mechanical resonance frequency of the concerned disk is attributed to volatile molecule adsorption (including water) as well as surface stress at the surface. Fig. 5 shows that each event is detected independently with our scheme. Moreover, it proves sensitive enough to detect air flow on neighboring resonators, with slight frequency changes ($\approx 3kHz$) of Disks 1 and 3 when the capillary is located above Disk 2. This is not the case for Disk 3 when the capillary is located above Disk 1 (and vice versa) as they are located further apart. This is to our knowledge the first demonstration of a capacity of spatial resolution in sensing with optomechanics. Note that the final mechanical frequencies, when the flow ends, do not reach back their initial values. We attribute this discrepancy to adsorbates remaining at the surface of the disk. This experiment demonstrates that individual and independent frequency variations of three optomechanical resonators can be measured simultaneously with a single optical channel.

## Discussion.

Increasing throughput and capture cross-section by operating a large number of nanomechanical resonators while maintaining their exquisite performance has been a long-standing challenge. We have proposed here a photonics-based architecture to multiplex optomechanical resonators. Three disk resonators have slightly different diameters, hence different resonance wavelengths and mechanical frequencies. They can be addressed by using one or the other degree of freedom, or both, providing much freedom in the design and in the implemented multiplexing/demultiplexing scheme. In our demonstration, we use three lasers with three different wavelengths to address all the disks in parallel, and recover information of their mechanical motion and frequency in the electrical domain. We have shown that our scheme does not degrade frequency stability, and that it is possible to monitor simultaneously and in real time the mechanical eigenfrequencies of all resonators. We have performed air flow sensing with the three resonators simultaneously and independently, with resonators in the very-high frequency (VHF) range and 1 ppm frequency stability.

Our scheme uses only one optical input/output and one photodiode. Distributed Feed Back lasers, commonly used in the telecom industry, can be used for the multiplexing, while multiple demodulation channels are used for demultiplexing. Considering that in standard large-scale photonics fabrication processes, 3-sigma variability in resonance wavelength is typically a few nm, and a free spectral range of ten's nm, we anticipate that several ten's resonators can be multiplexed with a single waveguide. Now-standard optical packaging enables connecting several hundreds of grating couplers with one fiber optic bundle. We anticipate that a few thousands resonators can then be multiplexed this way with our architecture.

This $10^3$ improvement in mechanical sensing data-rate has important implications for, for instance, chemical sensing: analysis time and instrument sensitivity can be improved by orders of magnitude. Many different resonators can be functionalized with layers specifically targeting different species, a key asset towards the realization of fast, high-resolution electronic noses and odor recognition. This is also the case for biomarker pattern recognition[6] with arrays of functionalized optomechanical microdisks in microfluidic channels. Such high-frequency devices have demonstrated their capability to retain high displacement sensitivity in liquid[30,31]. Our demonstration was performed with VHF resonators, and GHz frequencies have become commonplace for optomechanical devices. This translates into one to two orders of magnitude improvement in mechanical bandwidth compared to electrical resonators in the range of 10-100 MHz[4]. This is crucial for applications where abrupt events have to be detected at high speed such as mass spectrometry. With arrays of 10's electrical resonators, systems are today limited to analysis of a few hours. With the correct instrumentation and arrays of thousands of GHz optomechanical resonators, similar analysis can be performed in minutes with trace amount sensitivity.

## Methods:

**Fabrication process:**

The optomechanical detectors were realized with a Very Large Scale Integration fabrication process in an industrial-grade clean room, from 200 mm Silicon-on-Insulator wafers with 220 nm thick silicon top layer. A first lithography and partial etching (70 nm) of the grating couplers are performed. Then, waveguides and microdisks are patterned, using a lithography step with variable shape beam and dry etching. Silicon is then locally highly doped (boron, $5 \times 10^{19}$ at cm$^{-3}$) for electrical actuation while preserving good optical properties elsewhere. The metal lines (AlSi) are realized by depositing and patterning silicon oxide, depositing metal and a final planarization. Then a sacrificial layer (silicon oxide) is deposited, followed by a deposition of a ~200 nm-thick amorphous silicon layer, which is etched open only on top of the optomechanical resonators and grating couplers. A good alignment (below 100 nm) is critical for this step, as the aperture has to be located precisely on top of the platform. Finally, the microdisks are released by vapor HF etching.

**Set-up for multiplexing experiments**

The set-up used for multiplexing experiments is shown in Fig. 2. We generate the three input optical fields with three tunable external cavity lasers (Yenista Optics-EXFO). Each light field is then coupled to a polarization-maintaining fiber (Thorlabs) and is polarization controlled (Thorlabs). The three optical signals are then combined using a fiber optic coupler (Thorlabs), and are sent to the cavity optomechanical platform. The cleaved ends of the used monomode fibers are aligned with respect to the grating couplers with nanopositioning translations stages (Newport). The modulated optical signals are then measured with a photodiode (Newport). The generated voltage is measured with a digital lock-in amplifier (LIA) (Zurich Instruments). Inside the LIA, three demodulation channels operate simultaneously to demodulate the generated electrical signal at the microdisks mechanical resonance frequencies $f_{m1}$, $f_{m2}$ and $f_{m3}$, respectively. The demodulation bandwidth is set to 250 Hz for each demodulator, which is much smaller than the difference in frequency between two consecutive disks. In parallel three oscillators generate a voltage superposition, enabling to coherently drive the motion of the microdisks at their mechanical resonance frequencies (see Supplementary Fig. 12). In closed loop operation, the phase signal of each channel is the input signal of a phase locked loop (PLL) embedded in the LIA. The PLLs outputs are used to modify the frequencies of the actuation voltages generated by the oscillators..